# Transverse emittance measurements on an S-band photocathode rf electron gun *

**J.F. Schmerge, P.R. Bolton, J.E. Clendenin, F.-J. Decker, D.H. Dowell, S.M. Gierman, C.G. Limborg, B.F. Murphy**

Stanford Linear Accelerator Center, Menlo Park, CA 94025 USA

## Abstract

Proposed fourth generation light sources using SASE FELs to generate short pulse, coherent, X-rays require demonstration of high brightness electron sources. The Gun Test Facility (GTF) at SLAC was built to test high brightness sources for the proposed Linac Coherent Light Source at SLAC. The transverse emittance measurements are made at nearly 30 MeV by measuring the spot size on a YAG screen using the quadrupole scan technique. The emittance was measured to vary from 1 to 3.5 mm-mrad as the charge is increased from 50 to 350 pC using a laser pulse width of 2 ps FWHM. The measurements are in good agreement with simulation results using the LANL version of PARMELA.

*Contributed to*
*The 23rd International Free Electron Laser Conference*
*Darmstadt, Germany*
*20-24 August 2001*

* Work supported by Department of Energy contract DE–AC03–76SF00515.

# Transverse emittance measurements on an S-band photocathode rf electron gun

**J.F. Schmerge, P.R. Bolton, J.E. Clendenin, F.-J. Decker, D.H. Dowell, S.M. Gierman, C.G. Limborg, B.F. Murphy**

*SLAC, Menlo Park, CA 94025, USA*

*Corresponding author: J.F. Schmerge, SLAC, MS69, 2575 Sand Hill Rd, Menlo Park, CA 94025, USA., e-mail: Schmerge@slac.stanford.edu, Tel: 650-926-2320, Fax: 650-926-4100.*

Abstract

Proposed fourth generation light sources using SASE FELs to generate short pulse, coherent, X-rays require demonstration of high brightness electron sources. The Gun Test Facility (GTF) at SLAC was built to test high brightness sources for the proposed Linac Coherent Light Source at SLAC. The transverse emittance measurements are made at nearly 30 MeV by measuring the spot size on a YAG screen using the quadrupole scan technique. The emittance was measured to vary from 1 to 3.5 mm-mrad as the charge is increased from 50 to 350 pC using a laser pulse width of 2 ps FWHM. The measurements are in good agreement with simulation results using the LANL version of PARMELA.

PACS Codes: 29.25.Bx, 29.27.Ac, 41.60.Cr, 41.85.Ar

*Keywords: Photoinjectors; Low Emittance; Free Electron Lasers*

## 1. Introduction

The Gun Test Facility (GTF) is dedicated to the development of a high brightness electron source capable of producing a 1 mm-mrad emittance beam with at least 1 nC of charge to drive the Linac Coherent Light Source (LCLS) [1]. PARMELA simulations indicate the possibility of the GTF producing the desired beam using an emittance compensated 1.6 cell S-band photocathode gun with an approximately flat-top temporally shaped laser pulse driving the cathode [2]. Emittances measured at low charge (< 350 pC) with relatively short pulse lengths and Gaussian pulse shapes will be presented here.

The results are in agreement with PARMELA simulations. This agreement suggests that PARMELA can be used to verify the new LCLS photoinjector design [3].

## 2. Experimental Setup

The GTF laser is an Nd:glass based system utilizing chirped pulse amplification. After the amplifier is an optical compressor with one of the compressor gratings mounted on an optical rail to allow for pulse length adjustments by varying the grating separation and thus the compression ratio. The resulting IR pulse is then frequency doubled and quadrupled in a 1.25 mm long, type I, BBO crystal and a 0.5 mm long, type I BBO crystal respectively. A streak camera with 1 ps resolution in the UV was used to make temporal pulse shape measurements. Typical UV pulse length measurements for two different compression settings are plotted in Figure 1. Averaged over several measurements the shapes are near Gaussian with FWHM of 1.8 and 4.3 ps.

The RF gun installed at the GTF is a 1.6 cell, S-band, photocathode, rf gun with cavity symmetrization to remove dipole fields and the associated emittance growth [4]. The current cathode is a 1 cm diameter piece of single crystal copper (100 orientation) brazed onto the polycrystalline copper backplate. A 20 cm long emittance compensating solenoid [5] is located at the gun exit followed by a single 3 m traveling wave SLAC linac structure. The distance from the cathode to the input of the linac section is approximately 90 cm.

The transverse emittance of the e-beam is measured downstream of the linac booster by measuring the beam size on a 20 mm diameter, 0.5 mm thick YAG screen [6] as a function of quadrupole current (quadrupole scan technique). The YAG screen is 1.2 m downstream of the last quadrupole and is installed at normal incidence to the e-beam.

The light is extracted through a standard viewport and sent to an analog CCD camera. The resolution of the optical imaging system was measured to be approximately 50 μm. Images are captured and stored with an 8-bit frame grabber.

At each quadrupole current five images are captured along with one background image. After background subtraction the images are projected onto the horizontal axis, and the wings truncated at the point that is 5% of the peak value. A true rms calculation is performed on the truncated projection to find the rms beam size. The beam widths are averaged over the five data points and the emittance fit uses weighting inversely proportional to the rms spread in the data points. The emittance fit is calculated using the standard linear matrix along the beam line [7] and thus does not include space charge along the drift region between quadrupole and screen. A typical quadrupole scan is shown in Figure 2 with the measured rms beam sizes as a function of the quadrupole strength plotted as diamonds.

## 3. Results

We have measured the emittance as a function of charge for the two laser pulse lengths shown in Figure 1. Figure 3 plots as discreet points the minimum rms normalized emittances measured at a given charge for both the 1.8 and 4.3 ps FWHM laser pulses. The solenoid field has been optimized for each data point plotted. Due to laser intensity fluctuations, the rms charge fluctuation over each measurement is approximately 15%. The accelerator parameters are as follows; gun gradient of 110 MV/m, extraction phase of 38°, linac gradient of 8.5 MV/m and the linac phase which produces minimum energy spread. The minimum energy spread for the 4.3 ps pulses occurs at a linac phase of -13° and for the 1.8 ps data it is -20° relative to the crest of the wave. The minimum measured emittance is just under 1 mm-mrad for a 1.8 ps pulse at a charge of 105 pC and the maximum emittance is 3.5 mm-mrad at 350 pC. The uncertainty in the emittance measurements is dominated by charge fluctuations. Each point on the emittance curve can be reproduced to the order of 20%.

A series of simulations with the Los Alamos National Laboratory version 3 of PARMELA were conducted to compare with the measured results. Figure 3 shows the simulated normalized rms emittance as a function of charge for two different laser pulse lengths. These lines are within the range of the measurements presented here up to 250 pC.

As the charge is increased above 50 pC, PARMELA predicts the 4 ps data to have higher emittance than the 2 ps data despite the decrease in the space charge force for the longer pulse. The increase in emittance is due to a mismatch of the beam to the linac booster gradient. A proper match is achieved for laminar flow when equation 1 is satisfied, where $\gamma$ is the lorentz factor, $I_p$ is the peak current, $I_A$ is 17 kA, $\sigma_w$ is the rms beam size at a waist located at the linac entrance and $E_l$ is the linac gradient [8]. The transverse phase space at the quadrupole location of 5 central time slices are shown in Figure 4 for the 2 ps (top) and 4 ps (bottom) pulses for 300 pC from Figure 3. The graphs are labelled with the slope, s, of the linear regression, the emittance, $\varepsilon$, of the slice and the fraction of particles in that particular slice. For the 2 ps pulse, the 3 central slices, which contain 75% of the particles, are well aligned. For the 4 ps case the slice emittance is smaller, but the strong misalignment results in a larger projected emittance. When the linac gradient in the 4 ps case is reduced to satisfy equation 1, the simulated normalized rms emittance is also decreased.

$$E_l = \frac{1}{\sigma_w}\sqrt{\frac{I_p}{2I_A\gamma}} \qquad 1$$

## 4. Conclusions

We have measured the transverse emittance from an emittance compensated photoinjector at low charge using a quadrupole scan technique. The emittance was measured to vary from 1 to 3.5 mm-mrad as the charge is increased from 50 to 350 pC using a laser pulse width of 1.8 ps FWHM. The measurements are in reasonable agreement with PARMELA simulation results. In the future we will measure the emittance at higher charges and with temporally flat-top laser pulses.

*5 Acknowledgements*

*The Authors would like to thank D. Palmer from SLAC and J. Rosenzweig from UCLA for use of the single crystal copper cathode. We would like to acknowledge L. Serafini for interesting discussions about slice emittance and phase space alignment. We would also like to thank R. Ettelbrick of Positive Light for the loan of a Pockels cell to keep our laser operational. SLAC is operated by Stanford University for the Department of Energy under contract number DE-AC03-76SF00515.*

FIGURE 1

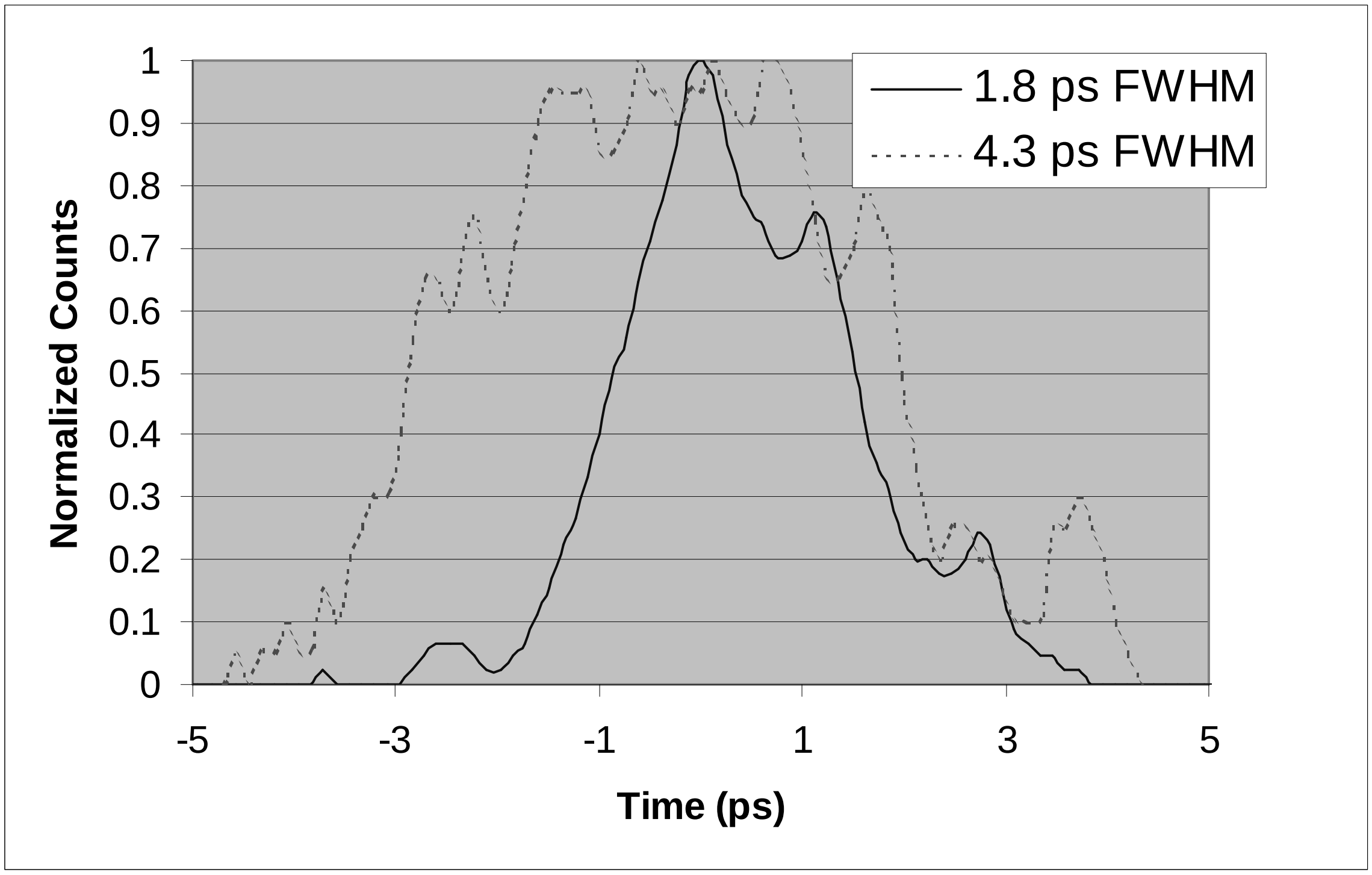

FIGURE 2

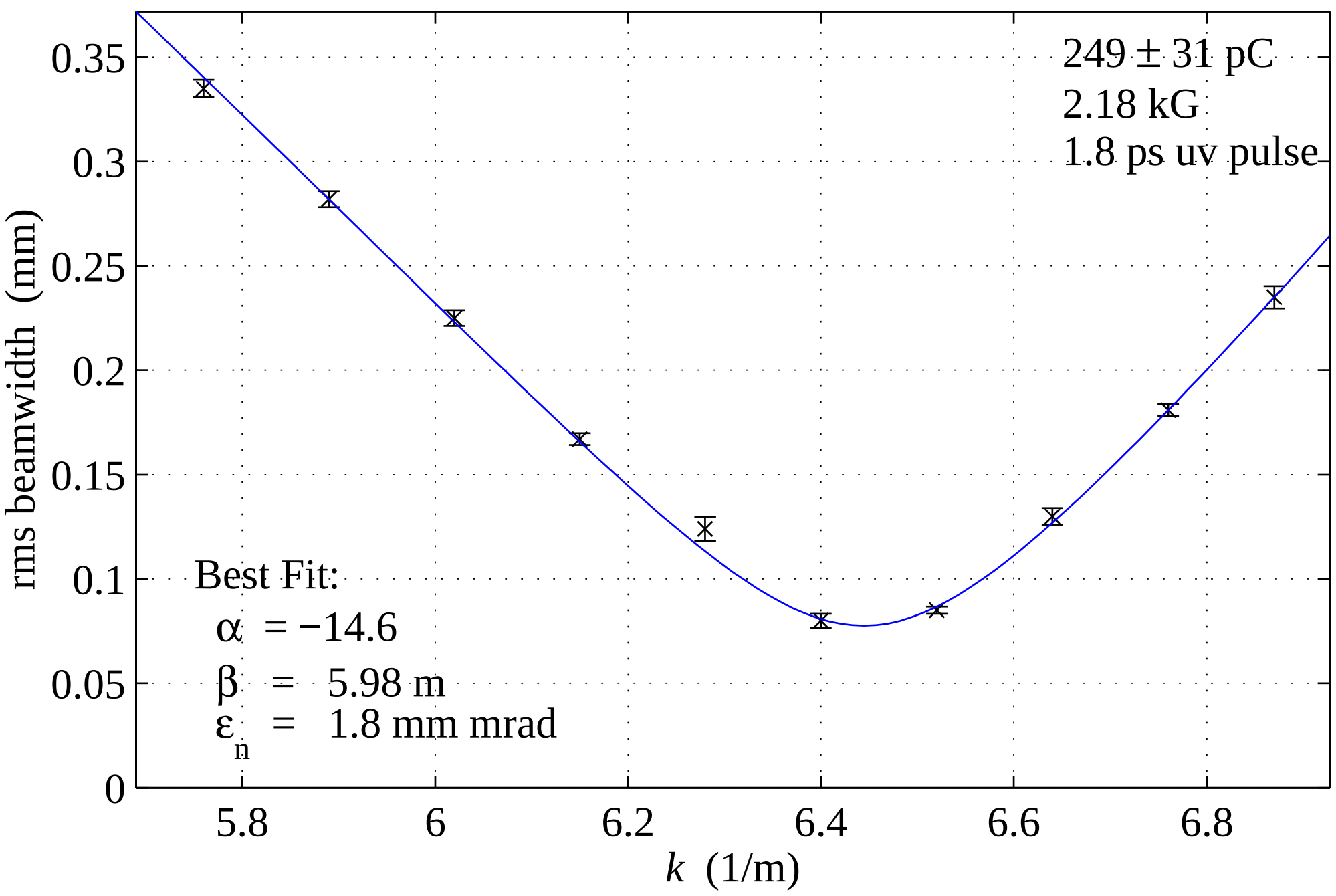

FIGURE 3

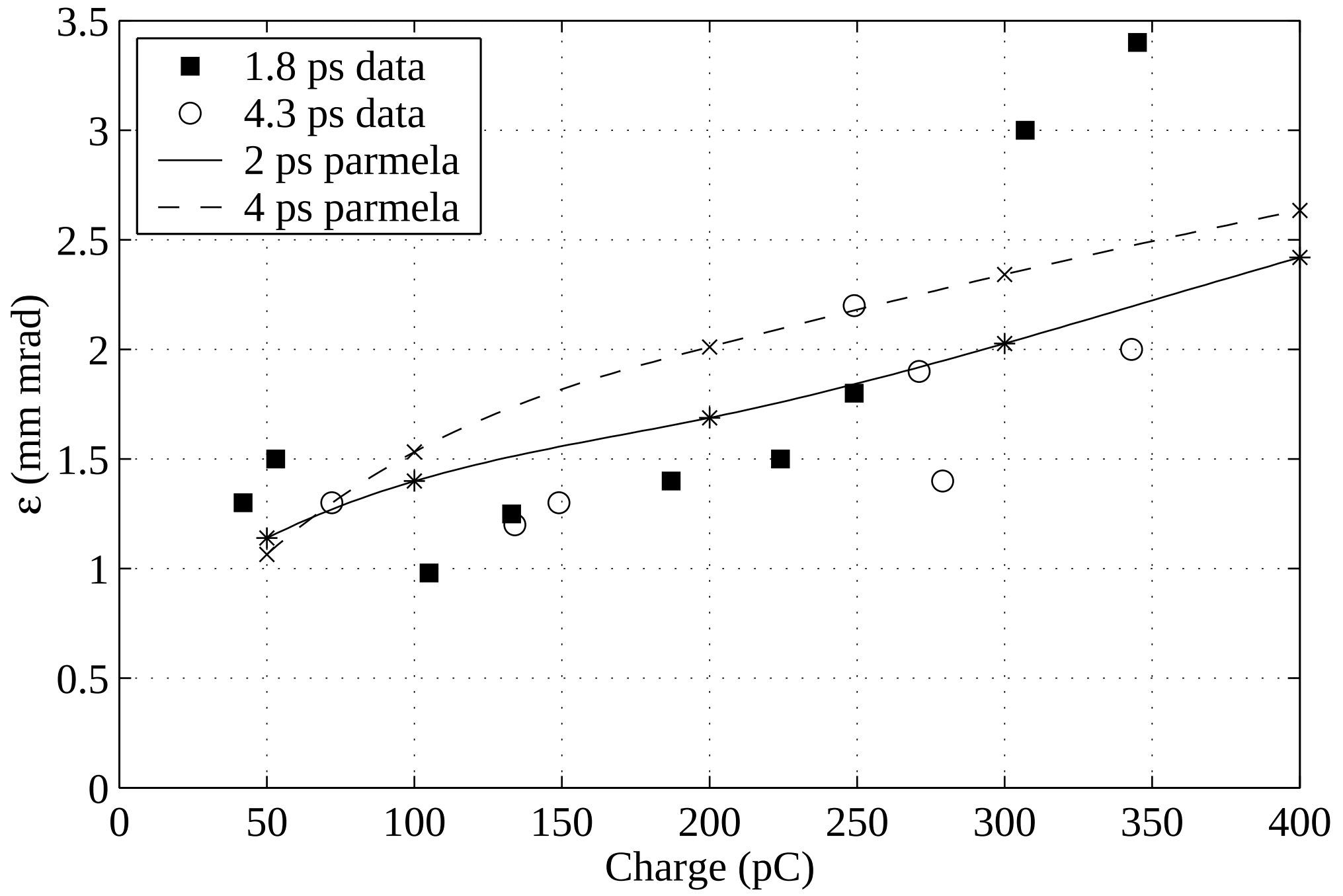

FIGURE 4

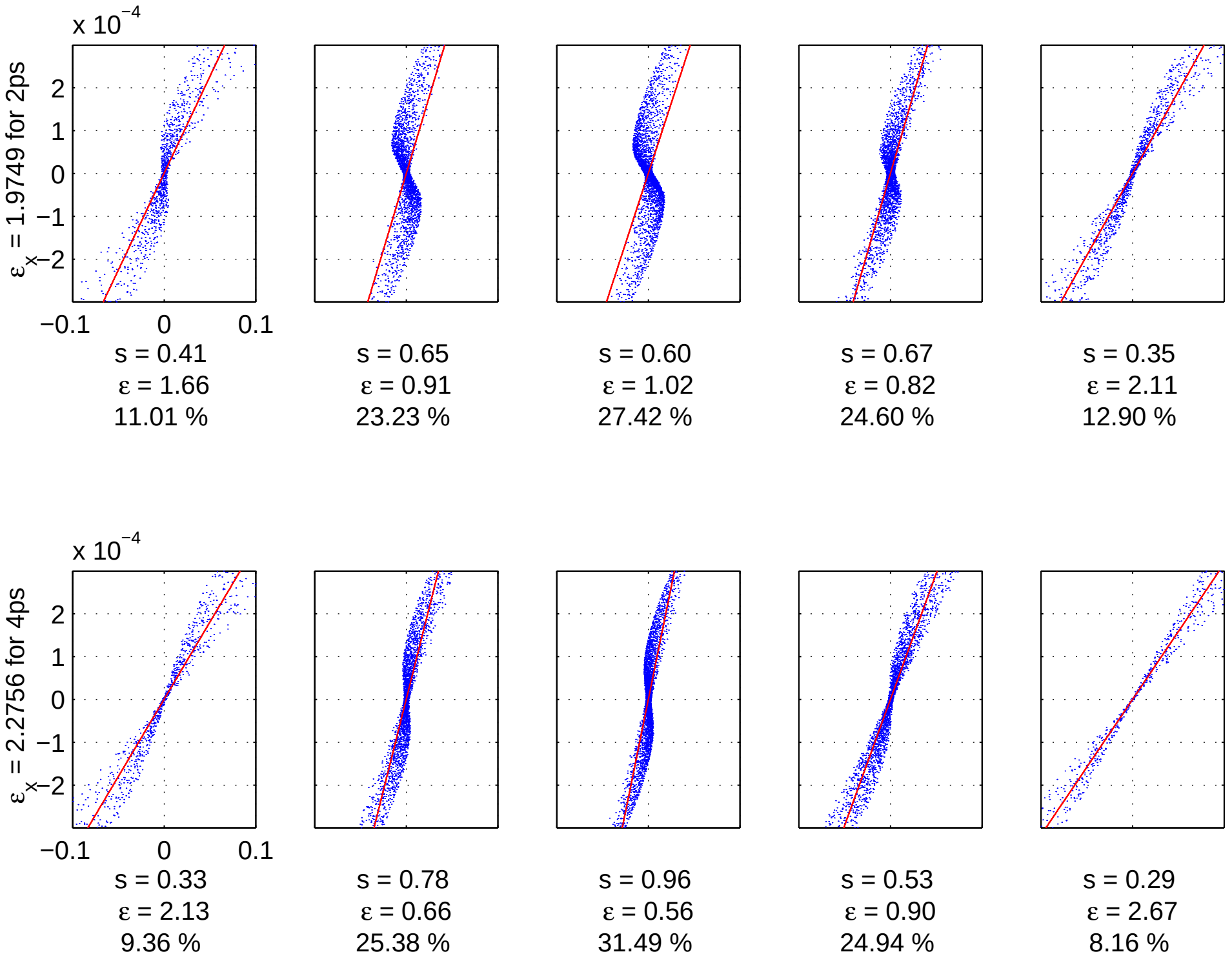

Figure Captions

Figure 1: The laser pulse shapes used in the experiment as measured with a 1 ps resolution streak camera. The average FWHM for the short and long pulse length is indicated in the figure.

Figure 2: The measured rms beam size as a function of quadrupole strength is plotted for a typical quadrupole scan. The plotted error bars are the standard deviation of the five measured beam sizes. The solid line is the fit with the Twiss parameters listed.

Figure 3: The minimum measured rms normalized emittance as a function of charge for two different laser pulse lengths are plotted as points. The lines are the minimum rms normalized emittance as determined by the simulation code PARMELA.

Figure 4: Transverse phase space plots for the 5 central time slices (out of a total of 7 evenly spaced slices) for the 2 ps case (upper) and 4 ps case (lower) for a charge of 300 pC. The abscissas are position in cm, the ordinates the corresponding divergence in radians. The slope, s, in $10^{-4}$-rad/cm, of the linear fit to the particles as well as the normalized rms emittance, $\varepsilon$, in mm-mrad, and percentage of charge in each slice is reported. The projected normalized rms emittance in mm-mrad for the total charge (7 slices) is given on the left. The slices in the 2 ps case are much better aligned leading to lower total projected emittance than the 4 ps case.

Endnotes